\documentclass[%
 reprint,
 amsmath,amssymb,
 aps,
]{revtex4-1}

\usepackage{graphicx}
\usepackage{dcolumn}
\usepackage{bm}

\begin{document}


\title{Active Rouse Chains}

\author{Dino Osmanovic}
\email{d.osmanovic@ucl.ac.uk}
\author{Yitzhak Rabin}%

\affiliation{Department of Physics, and Institute of Nanotechnology and Advanced Materials,
Bar-Ilan University, Ramat Gan 52900, Israel}

\date{\today}

\begin{abstract}
We consider how active forces modeled as non-thermal random noise affect the average dynamical properties of a Rouse polymer. As the power spectrum of the noise is not known we keep the analytical treatment as generic as possible and then present results for a few examples of active noise. We discuss the connection between our results and recent experimental studies of dynamics of labeled DNA telomeres in living cells, and propose new chromatin tracking experiments that will allow one to determine the statistical properties of the active forces associated with chromatin remodeling processes.

\end{abstract}

\pacs{Valid PACS appear here}
\maketitle


\section{Introduction}
The dynamics of polymer chains has long been of interest. From a purely physical standpoint, polymers display interesting behaviour in that they are viscoelastic, leading to complicated dynamics under stress. From a practical consideration, polymer chains are ubiquitous in biological systems, in addition to their wide industrial applications. The simplest theory of polymer dynamics, the Rouse model\cite{doiandedwards}, has been well studied and explains well the dynamical properties of polymers when hydrodynamic interactions and entanglements are not important.

More recently, there has been interest in \textit{active} systems. These systems are inherently non-equilibrium, consuming energy to produce motion. Examples include swimming bacteria or flocks of birds\cite{Ramaswamy:2010}. In the polymer context, many biopolymers are acted on by a variety of processes which consume energy (ATP), in order to remodel or translate the polymer\cite{Vignali:2000,Bruinsma:2014}. It is clear that these types of processes will not obey the  fluctuation-dissipation relation and, as such, the dynamics of active polymers may deviate markedly from standard Rouse model behavior. 

\begin{figure}[h!]
\begin{center}
\includegraphics[width=80mm]{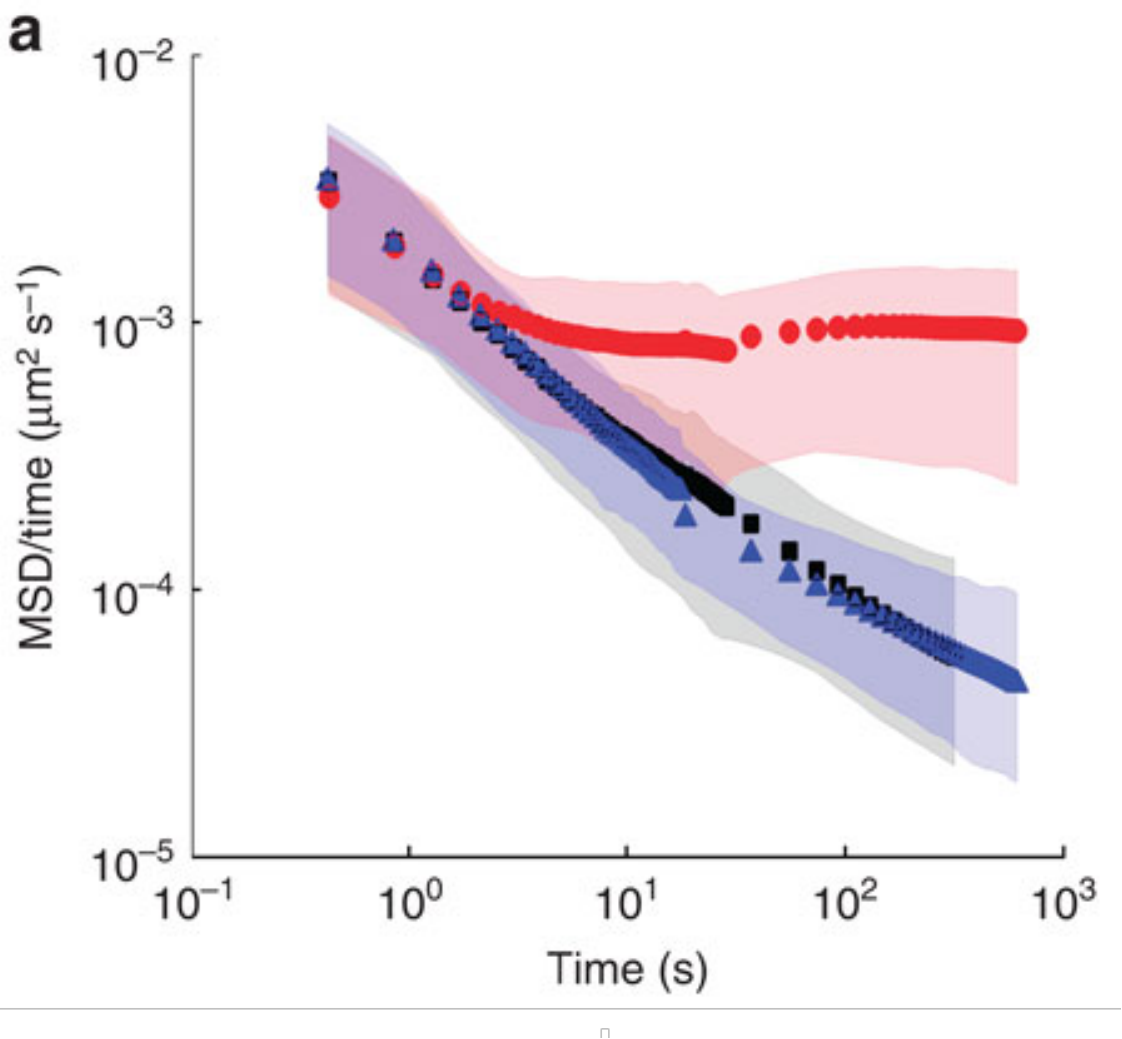}
\caption{Reproduced, with permission, from Bronstein \textit{et. al.}\cite{Bronstein:2015}, `` MSDs divided by time for telomeres in Lmna+/+ cells (black squares, N=474), Lmna−/− cells (red circles, N=503) and Lmna−/− cells after transfection with eGFP–pre-lamin A (blue triangles, N=220). Symbols designate the average locus MSD while shaded areas mark the s.d. of all single loci MSDs. " }
\label{fig:figexp} %
\end{center}
\end{figure}

Recent theoretical work in this area has considered systems which are driven by athermal noise in a viscoelastic medium\cite{Sakaue:2016,Vandebroek:2015}, in order to explain the observed deviations of the subdiffusive properties of chromatin from the predictions of the Rouse model\cite{barkai2012single,levi2005chromatin,bronstein2009transient,weber2012nonthermal}. Still, some intriguing questions remain unanswered. Even though the properties of the active noise are largely unknown and  the characterization of the power spectrum of active fluctuations is a topic of current interest\cite{Levine:2009,Bruinsma:2014}, the works of references  \cite{Sakaue:2016,Vandebroek:2015} did not address  systems with a generic noise (e.g., not necessarily with exponentially decaying temporal correlations), nor have correlations in active noise along the chain been considered. Furthermore, for what is perhaps the most important  biological example of a polymer being acted upon by active processes, chromatin, there are dynamical behaviors which are difficult to understand from the point of view of standard theories of polymer dynamics. For instance, Bronstein \textit{et. al.}\cite{Bronstein:2015} observed that when laminA is present, chromatin appears to follow standard polymer dynamics, albeit with a subdiffusion that is slower than predicted by the Rouse model. However, when laminA is removed, the motion of  labeled telomeres (and centromers) crosses over to normal diffusion orders of magnitude faster than in normal nuclei (see fig. \ref{fig:figexp}).

 In this paper, we consider a Rouse chain of frictional massless beads connected by linear (phantom) springs being acted upon by a generic active noise, and show how different kinds of noise can lead to quite different dynamical regimes. In particular, using this framework we try to understand the anomalous dynamical properties of chromatin. We then discuss the best quantities to measure in order to understand the statistical properties of the active noise in such biopolymer systems.

\section{Theory of Rouse Chain with arbitrary active forces}
We start with the standard Rouse dynamics equations for beads connected by elastic springs:
\begin{equation} \label{eq:rouse}
\gamma \frac{\partial \mathbf{R_n}(t)}{\partial t} = k \frac{\partial^2 \mathbf{R_n}(t)}{\partial n^2} + \mathbf{f_n}(t)
\end{equation}
where $\gamma$ is the damping constant, $k$ is the spring constant and $f_n$ is some random thermal force that acts on the nth bead, which has the following statistical properties:
\begin{align} \label{eq:thermal}
\langle f_{ni}(t)\rangle&=0 \\
\langle f_{ni}(t)f_{mj}(t')\rangle &=2 \gamma  k_B T \delta_{ij}\delta(n-m)\delta(t-t')
\end{align}
where $k_b$ is the Boltzmann constant and $T$ is the temperature. The labels $i$ and $j$ refer to the different components of the vector.

Following \cite{Vandebroek:2015}, we wish to introduce active forces to the system. These forces do not obey the fluctuation-dissipation theorem. We model them by adding an extra random force to eq. \ref{eq:rouse}
\begin{equation} \label{eq:activerouse}
\gamma \frac{\partial \mathbf{R_n}(t)}{\partial t} = k \frac{\partial^2 \mathbf{R_n}(t)}{\partial n^2} + \mathbf{f_n}(t) + \mathbf{A_n}(t)
\end{equation}
where the stochastic variable $\mathbf{A_n(t)}$ is the active force acting on monomer $n$ at time $t$. The only things we assume about the statistical properties of the noise is that it has zero mean, that the process is stationary and that the contributions to the correlation function are separable  in time and in monomer position.
 \begin{align} \label{eq:active}
\langle A_{ni}(t)\rangle&=0 \\
\langle A_{ni}(t)A_{mj}(t')\rangle &=\delta_{ij}B(t-t')C(n-m)
\end{align}
We transform this equation by taking the cosine transform with respect to $n$, and multiply the equation by $\frac{1}{N}\cos\left(\frac{p \pi n}{N}\right)$, where $N$ is the length of the chain. We then integrate with respect to $n$ between $0$ and $N$ and define the Rouse modes
\begin{equation} 
\mathbf{X_p}(t) =\frac{1}{N}\int_0^N \mathrm{d}n\cos\left(\frac{p \pi n}{N}\right)\mathbf{R_n}(t).
\end{equation}
This yields a set of uncoupled linear equations for the Rouse modes :
\begin{equation}
\gamma \frac{\partial X_p(t)}{\partial t} = -k_p X_p(t) + \widetilde{f}_p(t) + \widetilde{A}_p(t)
\end{equation}
where $k_p = \frac{3\pi^2 k_b T}{N b^2} p^2$ for all $p$. The transformed noises obey slightly different statistics. The thermal noise transforms to.
\begin{align} \label{eq:thermalmode}
\langle \widetilde{f}_{pi}(t)\rangle&=0 \\
\langle \widetilde{f}_{pi}(t)\widetilde{f}_{qj}(t')\rangle &=2 \frac{k_B T \gamma}{N} \delta{ij}\delta_{pq}\frac{1+\delta_{p0}}{2}\delta(t-t')
\end{align}
and the active noise transforms to:
 \begin{align} \label{eq:active}
\langle A_{pi}(t)\rangle&=0 \\
\langle \widetilde{A}_{pi}(t)\widetilde{A}_{qj}(t')\rangle &=\delta_{ij}B(t-t')I(p,q)
\end{align}
where the function $I(p,q)$ describes the coupling between the active forces acting on Rouse modes $p$ and $q$:
\begin{equation}
I(p,q)=\frac{1}{N^2}\int_0^N\mathrm\!\!\!{d}n\int_0^N\mathrm\!\!\!{d}m\, C(n-m)\cos\left(\frac{p \pi n}{N}\right)\cos\left(\frac{q \pi m}{N}\right)
\end{equation}
The solution of this equation is given by:
\begin{align}
&\mathbf{X}_p(t) = \mathbf{X}_p(0)\exp(-p^2 t/\tau_1)\\ \nonumber&+\frac{1}{\gamma}\int_0^t\exp(-p^2 (t-t')/\tau_1)(\mathbf{\widetilde{f}}_p(t')+\mathbf{\widetilde{A}}_p(t'))\mathrm{d}t'
\end{align}
where the polymer relaxation time $\tau_1$ is given by 
\begin{equation}
\tau_1=\frac{\gamma N^2 b^2}{3\pi^2 k_b T}. 
\end{equation}
Using the correlation functions of the modes $X_p(t)$ we can calculate the dynamical properties of the system. For example, the mean square displacement (MSD) of the center of mass is given by:
\begin{equation}
\langle R^2_{CM}(t)\rangle=\langle \left(\mathbf{X}_0(t)-\mathbf{X}_0(0)\right)^2\rangle
\end{equation}
and MSD of a labeled monomer $n$ is given by:
\small
\begin{align}
&\langle\left(R_n(t)-R_n(0)\right)^2\rangle\equiv g^{(n)}_1(t)=\langle R^2_{CM}(t) \rangle +\\ \nonumber & 4\sum_{p=1}^{\infty}\!\left(2\langle \mathbf{X_0}(t) \mathbf{X_p}(t)\rangle\!\!-\!\!\langle \mathbf{X_0}(0) \mathbf{X_p}(t)\rangle\!\!-\!\!\langle \mathbf{X_0}(t) \mathbf{X_p}(0)\rangle \right)\cos\left(\frac{p \pi n}{N}\right) \\ \nonumber &+8\sum_{p=1}^{\infty}\sum_{q=1}^{\infty}\left(\langle \mathbf{X_p}(t)\mathbf{X_q}(t)\rangle \!\!-\!\!\langle \mathbf{X_p}(t)\mathbf{X_q}(0)\rangle\right)\cos\left(\frac{p \pi n}{N}\right)\cos\left(\frac{q \pi n}{N}\right)
\end{align}
\normalsize
Once we know the correlation functions of the modes, many average dynamical quantities can be calculated.

For the generic correlation given by Eq. (\ref{eq:active}), the average motion of a single labeled monomer is given by (see S.I for derivation):
\begin{align}
&g^{(n)}_1(t)=\frac{6 k_b T}{N \gamma} t+\frac{1}{\gamma^2}I(0,0)\!\!\int_0^t\!\!\!\!\mathrm{d}a B(a)(t-a) + \\ \nonumber
&\sum_{p=1}^{\infty} \frac{12 k_b T \tau_1}{ N \gamma p^2}\left(1-\exp\left(-\frac{p^2 t}{\tau_1}\right)\right)\cos^2\left(\frac{p \pi n}{N}\right)+ \\ \nonumber &\sum_{p=1}^{\infty}\frac{8 \tau_1 I(0,p)}{p^2 \gamma^2}\Bigg[\left(1-\cosh\left(\frac{p^2 t}{\tau_1}\right)\right)\int_0^{\infty}\!\!\!\!\!\!\!\mathrm{d}a\,B(a)\exp\left(\frac{-a p^2}{\tau_1}\right)\\ \nonumber &+\int_0^t\!\!\! \mathrm{d}a\sinh\left(\frac{p^2(t-a)}{\tau_1}\right)B(a)\Bigg]\cos\left(\frac{p \pi n}{N}\right)+\\ \nonumber  &\sum_{p,q=1}^{\infty}\frac{8 \tau_1 I(p,q)}{\gamma^2(p^2+q^2)}\bigg[\int_0^{\infty}\!\!\!\!\!\mathrm{d}a B(a)\!\!\left(\!\!(1\text{-}e^{\text{-}\frac{p^2 t}{\tau_1}})e^{\text{-}\frac{a p^2}{\tau_1}}\!+\!(1\text{-}e^{\frac{q^2 t}{\tau_1}})e^{\text{-}\frac{a q^2}{\tau_1}}\right) \\ \nonumber&+2\int_0^t\!\!\!\!\mathrm{d}a B(a)\sinh\left(\frac{(p^2+q^2)(t-a)}{2 \tau_1}\right)\exp\left(\frac{(q^2-p^2)(t-a)}{2\tau_1}\right)\bigg]\\ \nonumber &\cos\left(\frac{p \pi n}{N}\right)\cos\left(\frac{q \pi n}{N}\right)
\end{align}

In the case where there are no correlations along the contour of the chain this expression is simplified further: using $I(p,q) = \frac{1}{N}\frac{1+\delta_{p0}}{2}\delta_{pq}$, we are left with
\begin{align} \label{eq:MSDnolength}
&g^{(n)}_1(t)=\frac{6 k_b T}{N \gamma} t+\frac{1}{N\gamma^2}\!\!\int_0^t\!\!\!\!\mathrm{d}a B(a)(t-a) + \\ \nonumber
&\sum_{p=1}^{\infty} \frac{12 k_b T \tau_1}{ N \gamma p^2}\left(1-\exp\left(-\frac{p^2 t}{\tau_1}\right)\right)\cos^2\left(\frac{p \pi n}{N}\right)+ \\ \nonumber  &\sum_{p=1}^{\infty}\frac{4 \tau_1}{N\gamma^2 p^2}
\int_0^{\infty}\!\!\!\!\!\mathrm{d}a B(a) \left(e^{-a p^2/\tau_1}- \frac{1}{2}\left(e^{-\frac{p^2}{\tau_1}|t-a|}+e^{-\frac{p^2}{\tau_1}|t+a|}\right)\right) \\ \nonumber &\cos^2\left(\frac{p \pi n}{N}\right)
\end{align}
Eq. \ref{eq:MSDnolength} is a generic equation for the MSD of a labeled particle as a functional of the correlation function in time $B$ where the noise is not correlated along the chain. It can be seen that the MSD is just a sum of two terms, one corresponding to the normal Rouse contribution, with the extra term resulting from the active noise.

At any particular time, the scaling of the MSD with $t$, (i.e, $\approx t^{\alpha}$) can be calculated by:
{\bf why is the argument t' rather than simply t?}
\begin{equation} \label{eq:scaling}
\alpha(t')=\frac{\partial \log(g^{(n)}_1(t))}{\partial \log(t)}\bigg|_{t=t'}
\end{equation}
which is a functional of the correlation function $B$. There are only a few physical limitations on the form of $B$. Firstly, $\int_0^\infty \mathrm{d}t B(t)$ is positive. Secondly, $B(t)$ is positive as $t\to0$ as the noise is perfectly correlated with itself at the same time. Other than these constraints, we are free to choose whichever form of $B$ we like. For most systems, the precise form of the power spectral density of the active noise is unknown.

\section{Results}
Different forms of the correlation function can lead to quite different dynamical regimes. For instance, if the active noise is assumed to be uncorrelated at different points along the chain and at different times:
 \begin{equation}
\langle A_{ni}(t)A_{mj}(t')\rangle=2F\delta_{ij}\delta(t-t')\delta(n-m),
\end{equation}
where $F$ is a constant, one recovers the standard Rouse results, with a renormalized temperature $T^{(\text{active})}=T + F/k_b\gamma$. This is reminiscent of approaches which treat activity as a higher effective temperature\cite{ganai2014chromosome,ghosh2014dynamics,grosberg2015nonequilibrium}.

\begin{figure}
\begin{center}
\includegraphics[width=80mm]{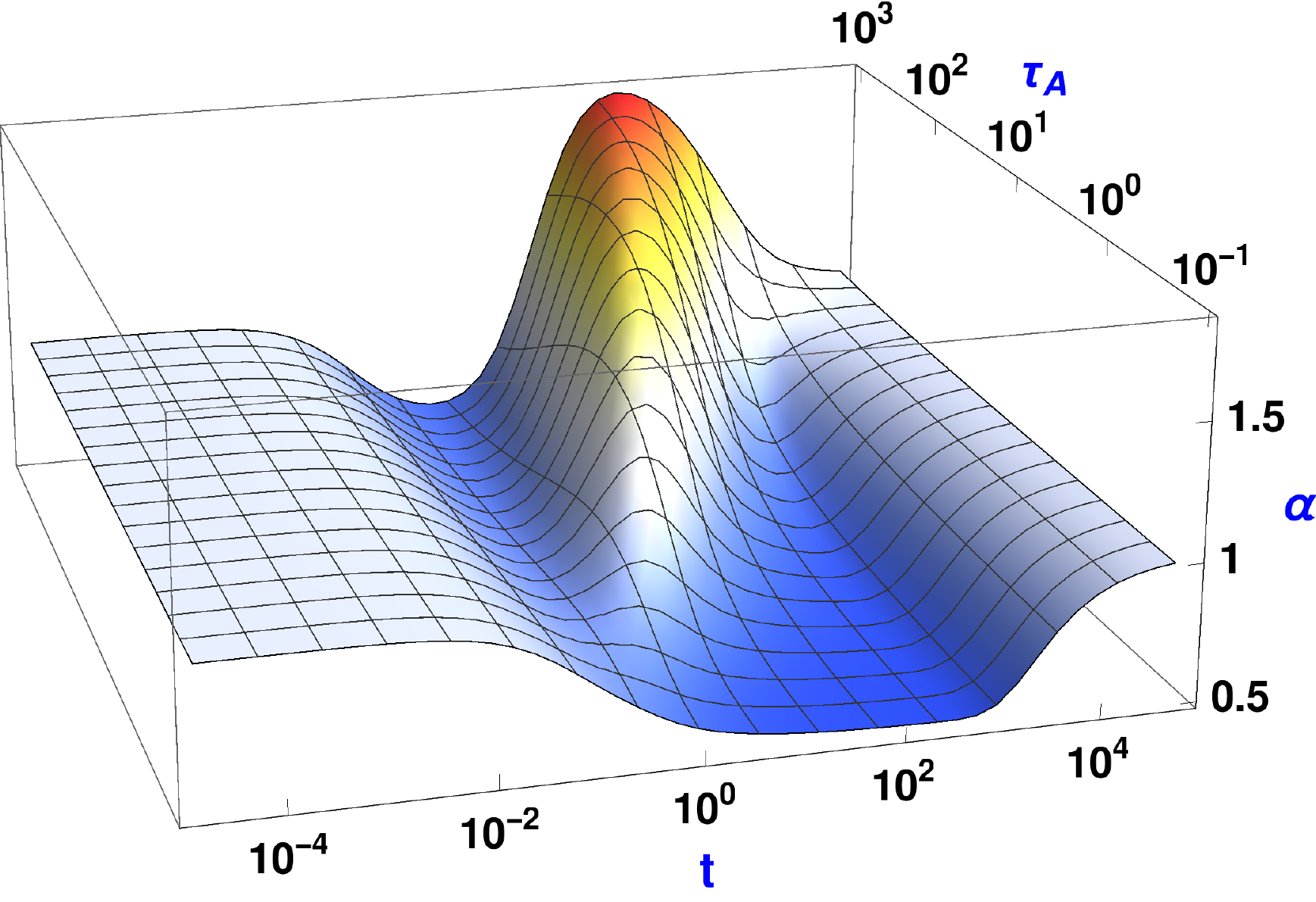}
\caption{The scaling exponent of the middle monomer as a function of both $t$ and the timescale of the exponential correlations $\tau_A$. The magnitude of the thermal and active forces is the same, the chain is 200 beads long and $\gamma=k=k_BT=1$.}
\label{fig:fig1} %
\end{center}
\end{figure}

A common assumption about active correlations is that they are exponentially correlated in time with no correlations along the chain (see e.g \cite{Sakaue:2016,Vandebroek:2015}):
\begin{equation}
B(t)=6F e^{-t/\tau_A}.
\end{equation} 
The integrals in eq. (\ref{eq:MSDnolength}) can be easily evaluated and the monomer MSD can be obtained:
\small
\begin{align}
&g^{(n)}_1(t)=\frac{6 k_b T}{N \gamma} t+\frac{6 F}{N\gamma^2}\tau_A \left(\tau_A \left(e^{-\frac{t}{\tau_A}}-1\right)+t\right) + \\ \nonumber &\sum_{p=1}^{\infty} \frac{12 k_b T \tau_1}{ N \gamma p^2}\left(1-\exp\left(-\frac{p^2 t}{\tau_1}\right)\right)\cos^2\left(\frac{p \pi n}{N}\right)+ \\ \nonumber  
&\sum_{p=1}^{\infty}\frac{24 F \tau_1 \tau_A}{N \gamma^2} \frac{\left(1-e^{-\frac{p^2 t}{\tau_1}}\right)+\frac{p^2 \tau_A}{\tau_1} \left(e^{-\frac{t}{\tau_A}}-1\right)}{p^2(1-p^2 \tau_A/\tau_1)(1+p^2 \tau_A/\tau_1)}\cos^2\left(\frac{p \pi n}{N}\right)
\end{align}
\normalsize
The consequences of such a noise can be calculated from the expression above. In particular, it is instructive to plot how the exponent $\alpha(t)$ (defined in Eq. (\ref{eq:scaling})) changes as a function of both $t$ and $\tau_A$ (see fig. \ref{fig:fig1}).
As is clear from fig. \ref{fig:fig1}, introducing correlations over some timescale $\tau_A$ can lead to significant deviations in dynamical behaviour from the standard Rouse case. When $\tau_A$ is small, the behavior does not deviate markedly from the Rouse model predictions; there is a very short initial period of normal diffusion with $\alpha=1$, followed by a long period of subdiffusive motion with $\alpha\approx 0.5$ (where the motion of the monomer is hindered by its coupling to the rest of the polymer chain) and finally crossover to normal diffusion (with the center of mass of the entire polymer) as the time approaches the longest polymer relaxation time $\tau_1$.

As $\tau_A$ is increased, the active noise is correlated over longer times, and superdiffusive behaviour ($\alpha>1$) can be observed. The larger the value of $\tau_A$, the larger this superdiffusive peak is. The period where the motion is superdiffusive is roughly when $t=\tau_A$, reminiscent of adding a resonance peak at this time. We analyze exponential noise in more detail in the SI.

When we add correlations along the chain we change some of the dynamical properties, but the results look qualitatively similar, albeit with somewhat larger exponents  $\alpha(t)$ (see SI). 

In the context of chromatin, such a noise cannot effectively describe the dynamics observed under some experimental conditions, such as the removal of laminA. A possible reason for the discrepancy is that while exponentially correlated noise is associated with a single timescale $\tau_A$, there is a broad range of processes which operate on chromatin over many different time scales. 

As an example, let us assume that the correlation function of the system is given as follows:
\begin{equation}
B(t)=\frac{1}{\log(\tau_b)-\log(\tau_s)}\int_{\tau_s}^{\tau_b} \frac{1}{\tau}\exp\left(-\frac{t}{\tau}\right)\mathrm{d}\tau
\end{equation}
This integral can be evaluated analytically, yielding:
\begin{equation}
\frac{\Gamma(0,t/\tau_b)-\Gamma(0,t/\tau_s)}{\log(\tau_b)-\log(\tau_s)}
\end{equation}
where $\Gamma$ is the incomplete gamma function. The power spectral density $F(\omega)$ of such a correlation behaves as a $\text{constant}$ when $\omega<1/\tau_b$, as $\sim1/\omega$ when $1/\tau_b<\omega<1/\tau_s$ (the celebrated $1/f$ noise\cite{1/f:1981}) and as $\sim1/\omega^2$ when $\omega>1/\tau_s$. Inserting this expression into eq. (\ref{eq:MSDnolength}) can tell us what the dynamics of a system under such noise would be.

\begin{figure}
\begin{center}
\includegraphics[width=80mm]{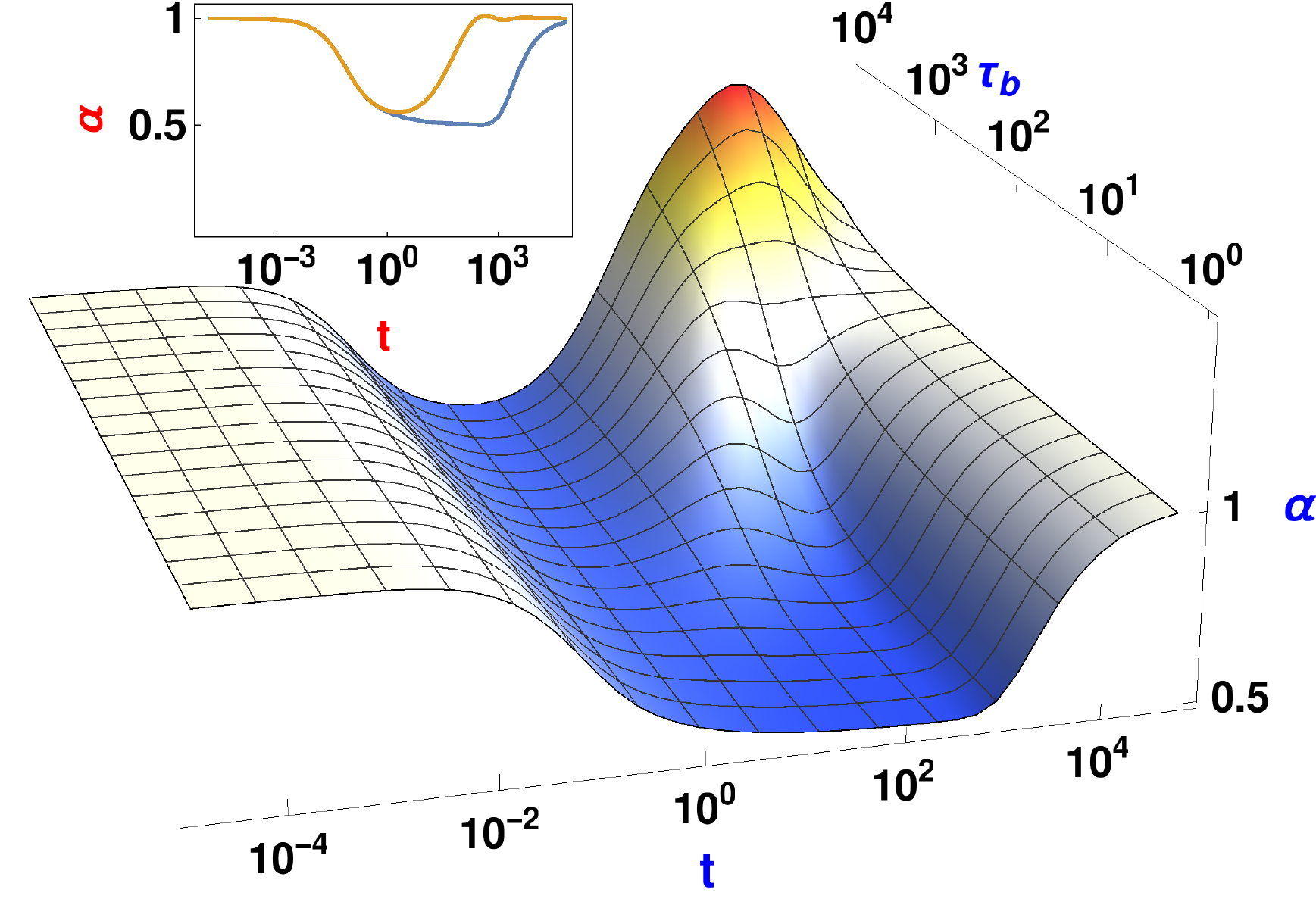}
\caption{The scaling exponent of the middle monomer, of a chain of 200 monomers, as a function of both $t$ and the longest correlation time $t_b$. In this case the amplitude of the active noise is only 7\% of that of the thermal noise and $\tau_{s}=10^{-3}$ (shorter than the monomer relaxation time). Inset: slices along $t_b\to 0$, the normal Rouse case, and $t_b=10^3$; activity over many timescales speeds up the transition to normal diffusion.}
\label{fig:fig2} %
\end{center}
\end{figure}

From fig. \ref{fig:fig2} it can be seen that having active noise correlations over many time scales narrows the time interval in which subdiffusion is observed. In particular, when $\tau_b$ approaches the longest relaxation time of the polymer, we can reproduce the qualitative behavior observed in the experiment by Bronstein \textit{et. al.} in the absence of laminA\cite{Bronstein:2015}, in that the transition to normal diffusion of a labeled monomer takes place several orders of magnitude faster than predicted by the standard Rouse model (see inset in fig. \ref{fig:fig2}).

\section{Discussion}
In our model, the polymer chain moves freely in the surrounding solvent and is always being acted upon by a variety of active processes. The direct application of the present model to chromatin in normal cells is hindered by several factors the most important of which is the presence of constraints that include the coupling of chromatin to the nuclear envelope in which laminA plays a major role\cite{Dechat:2010,Melcer:2012,DeVos:2010,Mahen:2013,Moir:2000}, and possibly also the cross-linking of chromatin by laminA that suppresses the motion of telomeres  on length scales $\gg 100 nm$, as suggested by \cite{Bronstein:2015}. According to this reasoning,  in the absence of laminA these constraints are released and resulting large-scale  motions of labeled telomeres are driven by a combination of thermal and active forces that can be qualitatively described by our model.

\begin{figure}
\begin{center}
\includegraphics[width=80mm]{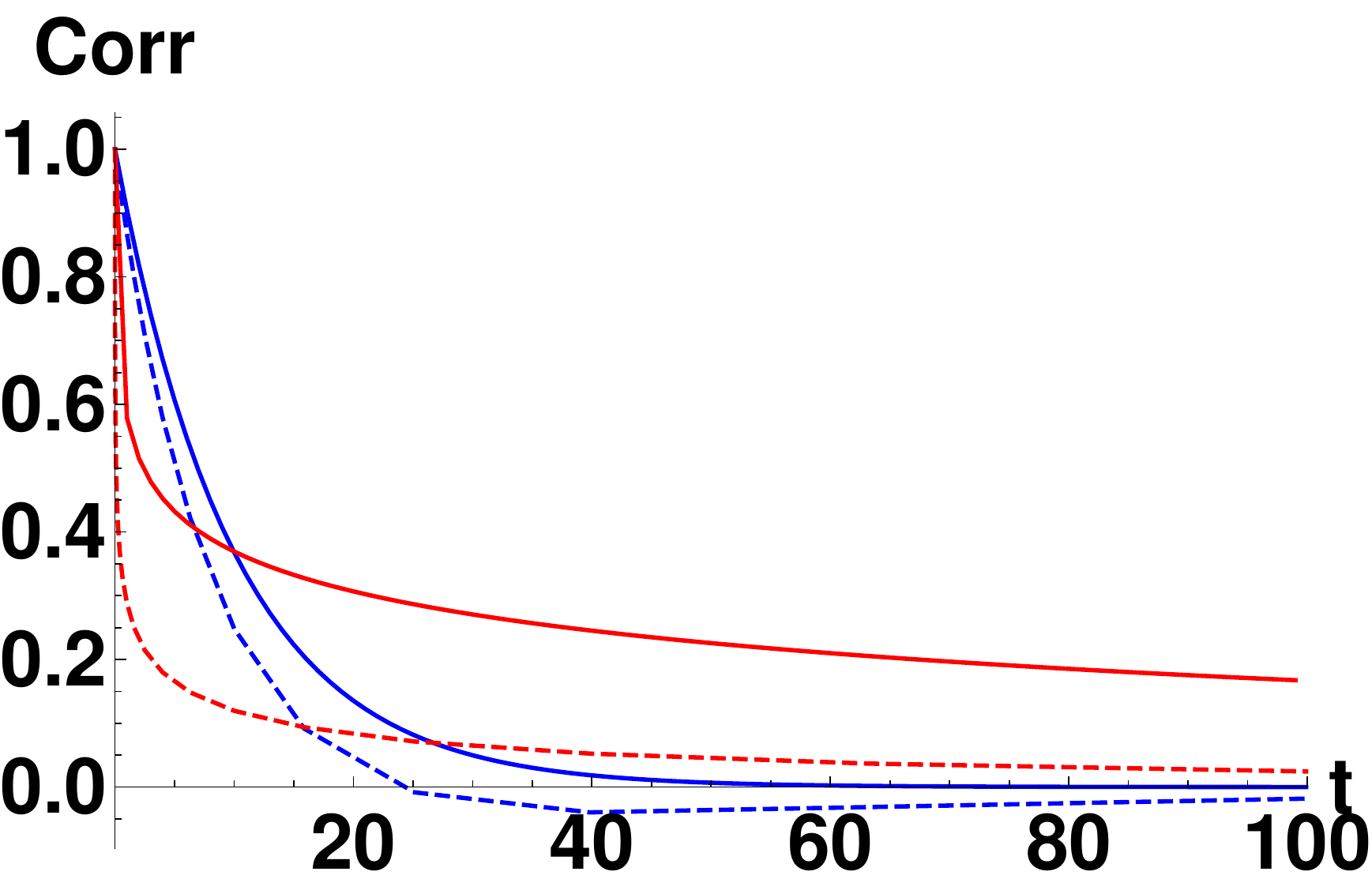}
\caption{The correlation (ignoring the delta function peak due to the thermal contribution) of the active noise (solid lines) and the velocity autocorrelation (dashed lines) for a system with exponentially correlated noise (blue) with a timescale of $\tau_A=10$ and a system with the $\approx 1/f$ noise described in text (red)}
\label{fig:fig3} %
\end{center}
\end{figure}

Lets assume that there exist experimental conditions in which the present model can be applied to chromatin. So far we considered the question of what kind of dynamics (MSD of labeled monomers) is expected for a given form of active forces that act on chromatin. One can also whether it is possible, by monitoring the trajectories of labeled monomers, to extract the statistical properties of the ATP-dependent forces that act on the chromatin. We now demonstrate that this is indeed feasible by measuring velocity correlations of labeled chain monomers. The solution of the Rouse equation is given, in terms of modes by:
\begin{equation}
\mathbf{R_n}(t)=\mathbf{X_0}(t)+2\sum_{p=1}^N \mathbf{X_p}(t)\cos\left(\frac{p\pi n}{N}\right)
\end{equation}
Differentiating with respect to time yields:
\begin{equation}
\mathbf{\dot{R}_n}(t)=\mathbf{\dot{X}_0}(t)+2\sum_{p=1}^N \mathbf{\dot{X}_p}(t)\cos\left(\frac{p\pi n}{N}\right)
\end{equation}
The time derivative of the mode can be separated into thermal and active contributions,
$\mathbf{\dot{X}_p}(t)=\mathbf{\dot{X}_{p,T}}(t)+\mathbf{\dot{X}_{p,A}}(t)$.  Therefore, for arbitrary temporal and contour correlations of the active force, the velocity correlation will be given by:
\begin{align}
\nonumber &\langle \mathbf{\dot{R}_n}(t)\mathbf{\dot{R}_m}(0)\rangle=\langle \mathbf{\dot{X}_{0,T}}(t) \mathbf{\dot{X}_{0,T}}(0) \rangle+\langle \mathbf{\dot{X}_{0,A}}(t) \mathbf{\dot{X}_{0,A}}(0) \rangle \\ \nonumber&+4\sum_{p=1}^{\infty}\langle \mathbf{\dot{X}_{p,T}}(t)\mathbf{\dot{X}_{p,T}}(0)\rangle \cos^2\left(\frac{p \pi n}{N}\right) \\ \nonumber &+2\sum_{p=1}^{\infty}\langle \mathbf{\dot{X}_{p,A}}(t)\mathbf{\dot{X}_{0,A}}(0)\rangle \cos\left(\frac{p \pi n}{N}\right)\\ \nonumber &+2\sum_{q=1}^{\infty}\langle \mathbf{\dot{X}_{q,A}}(t)\mathbf{\dot{X}_{0,A}}(0)\rangle \cos\left(\frac{q \pi m}{N}\right)\\ \nonumber &+4\sum_{p,q}^{\infty}\langle \mathbf{\dot{X}_{p,A}}(t)\mathbf{\dot{X}_{q,A}}(0)\rangle \cos\left(\frac{p \pi n}{N}\right)\cos\left(\frac{q \pi m}{N}\right)
\end{align}
We can see that the velocity correlation is given by the standard Rouse model with an active contribution. The velocity correlation of the standard Rouse model is a delta function in time, which arises due to the fact that inertia is neglected. In reality, there is some inertial timescale associated with the velocity correlation; however if this is smaller than the measurement time of any experimental realization of such a system it won't be seen. Therefore, in any system with active noise that is correlated on longer timescales or correlated along the chain, this should be visible from the velocity correlation function.

From fig. \ref{fig:fig3} it can be seen that the velocity correlation looks very similar to the active force correlation function with some minor differences. For instance, when the active noise is exponentially correlated, the velocity correlation can become negative for $t>\tau_A$, an indication that the force applied to the monomer leads to an increase in tension along the backbone of the polymer which eventually reverses the direction of the velocity after the force correlations have decayed. For the other noise we considered ($1/f$ noise), force correlations decay sufficiently slowly and negative velocity correlations are not expected. It is interesting to note that the velocity autocorrelation function declines faster than the active force correlations. This is due to the fact that the response of the chain is determined not only by the active force but also by the elastic force of the springs that opposes the active force. If one considers the center of mass velocity (there is no elastic force that acts on the center of mass), the velocity autocorrelation and that of the average force will be the same.

\section{Conclusion}
We calculated analytically the MSD of a labeled monomer of a Rouse polymer chain subjected to a combination of stochastic forces of thermal and active origins, for arbitrary temporal and contour separation correlations of the active force. We found that when the temporal correlations are characterized by a single time scale $\tau_A$, there is a transition from normal Rouse behavior (at a higher effective temperature) to superdiffusive motion at intermediate time scales, with increasing $\tau_A$. Curvilinear long-range motions of labeled chromatin have been observed experimentally in living cells (see ref. \cite{chuang2006long} ), but it is not clear at present whether the mechanism leading to these motions is connected to the superdiffusive regime predicted by our model. When multiple time scales are involved in the processes acting on the polymer ($1/f$ noise), one can observe an anomalously short subdiffusive regime followed by a transition to normal diffusion at times much shorter than the longest Rouse time of the polymer. The latter result provides a possible explanation for the recent observations of telomere dynamics in laminA-deficient cells \cite{Bronstein:2015}. We also propose a new method for studying the statistical properties of active forces in living cells by measuring the autocorrelation function of the velocity of labeled monomers.

\begin{acknowledgments}
This work was supported by the I-CORE Program of
the Planning and Budgeting committee and the Israel Science Foundation grant 1902/12. We would like to thank Yuval Garini for discussions.
\end{acknowledgments}
\bibliographystyle{unsrt}
\bibliography{ActivePolymer}
\end{document}